\documentclass[12pt]{article}

\catcode`\@=11
\def\marginnote#1{}
\newcount\hour
\newcount\minute
\newtoks\amorpm
\hour=\time\divide\hour by60
\minute=\time{\multiply\hour by60 \global\advance\minute
by-\hour}\edef\standardtime{{\ifnum\hour<12
\global\amorpm={am}%
        \else\global\amorpm={pm}\advance\hour by-12 \fi
        \ifnum\hour=0 \hour=12 \fi
        \number\hour:\ifnum\minute<10
0\fi\number\minute\the\amorpm}}
\edef\militarytime{\number\hour:\ifnum\minute<10
0\fi\number\minute}

\def\draftlabel#1{{\@bsphack\if@filesw {\let\thepage\relax
   \xdef\@gtempa{\write\@auxout{\string
      \newlabel{#1}{{\@currentlabel}{\thepage}}}}}\@gtempa
   \if@nobreak \ifvmode\nobreak\fi\fi\fi\@esphack}
        \gdef\@eqnlabel{#1}}
\def\@eqnlabel{}
\def\@vacuum{}
\def\draftmarginnote#1{\marginpar{\raggedright\scriptsize\tt#1}}
\def\draft{\oddsidemargin -.5truein
        \def\@oddfoot{\sl preliminary draft \hfil
        \rm\thepage\hfil\sl\today\quad\militarytime}
        \let\@evenfoot\@oddfoot \overfullrule 3pt
        \let\label=\draftlabel
        \let\marginnote=\draftmarginnote

\def\@eqnnum{(\theequation)\rlap{\kern\marginparsep\tt\@eqnlabel}%
\global\let\@eqnlabel\@vacuum}  }

\def\numberbysection{\@addtoreset{equation}{section}
        \def\theequation{\thesection.\arabic{equation}}}

\def\underline#1{\relax\ifmmode\@@underline#1\else
 $\@@underline{\hbox{#1}}$\relax\fi}

\catcode`@=12
\relax

\numberbysection
\def\beq{\begin{equation}}
\def\eeq{\end{equation}}
\def\bea{\begin{eqnarray}}
\def\eea{\end{eqnarray}}
\def\beqa{\begin{eqnarray}}
\def\eeqa{\end{eqnarray}}

\def\fin{\end{document}}
\begin{document}

\begin{titlepage}
\nopagebreak
\begin{flushright}
LPTENS-98/41\\
hep--th/9811108
\\
November  1998
\end{flushright}
\begin{center}
{\large
PROGRESS IN CLASSICALLY SOLVING\\
\medskip 
 TEN DIMENSIONAL SUPERSYMMETRIC\\ 
\medskip\smallskip  
REDUCED YANG--MILLS THEORIES}\\
{ Jean--Loup GERVAIS}\\
{\footnotesize Laboratoire de Physique Th\'eorique de l'\'Ecole Normale Sup\'erieure
\footnote{Unit\'e Propre du Centre National de la Recherche Scientifique, associ\'ee \`a l'\'Ecole
Normale Sup\'erieure et \`a l'Universit\'e de Paris-Sud.},\\ 24 rue Lhomond, 75231 Paris C\'EDEX
05, ~France.}\\ and { the late} \\ { Mikhail V. SAVELIEV
\footnote{ has been supported in
part by the Russian Foundation for Basic Research under grant
\# 98-01-00015 and by INTAS grant \# 96-690.}}\\
{\it (deceased on September 20th 1998) }\\
{\footnotesize{Institute for High Energy Physics, 142284, Protvino, Moscow region, Russia.}} 
\end{center}
\bigskip 
\begin{abstract}
It is shown that there exists an on--shell light cone gauge where 
half of the fermionic components of the super vector potential vanish,
so that part of the superspace flatness conditions becomes
linear. After reduction to $(1+1)$ space-time dimensions, the
general solution of this subset of equations is derived. The remaining non-linear
equations  are written in a  form which is analogous to  Yang
equations, albeit with superderivatives involving sixteen 
fermionic coordinates.
It is shown that this  non-linear part  may, nevertheless, 
 be solved by  methods similar to
powerful technics previously developed  for the
(purely bosonic) self--dual Yang Mills equations in four dimensions.  
\baselineskip .4 true cm
\noindent

\end{abstract}
\begin{center}
\textit{\bfseries TO THE MEMORY OF MIKHAIL SAVELIEV}
\end{center}
\textit{\bfseries 
This work was being completed when my very good friend and long time collaborator suddenly passed 
away. His family, his numerous friends and the whole scientific community have suffered an 
unbearable loss. It is my hope that this article, which describes the state of a part of  our
research program at the time of his death, will remain as a valuable tribute to his memory. }  
\begin{flushright}
 \textit{\bfseries  Jean-Loup Gervais}
\end{flushright}
\vfill
\end{titlepage}

\baselineskip .5 true cm

\section{Introduction} 
It has been  known already for more than ten years, see e.g. \cite{W86,AFJ88,CM89}, that ten
dimensional supersymmetric
Yang--Mills theories may be considered as integrable systems classically,  in a ``weak'' sense, 
since they admit  Lax--type representations in  superspace for the equations of motion.
The starting point in this direction  was the observation \cite{N81,W86,AFJ88} that the field
equations of  these theories are equivalent to the  constraint that the purely fermionic components
of the supercurvature vanish. So far, the existence of this Lax pair
has not been so useful 
however, since the role of  spectral parameter is played 
by  a light-light vector.  More recently, the interest was revived into  (suitably reduced) ten dimensional
supersymmetric Yang--Mills theories in the large
$N$ limit since they have been actively  considered in  the search for the M theory (see e.g
refs\cite{BFSS96,M97,DVV97}). This has motivated us to return to the use of the flatness condition
in superspace in order to derive non trivial classical solutions. One may hope, in particular that
the problem will be simpler after the reduction process, which we will simply perform by looking
for classical solutions that do not depend upon a certain set of space coordinates. 
We shall indeed make progress  after reducing to $1+1$ space--time
dimensions, which seems  to be 
the most natural  choice at the present time. 
 
Let us first recall some standard formulae in order to establish the notations. 
In ten dimensions the dynamics is specified by the standard action 
\beq
S=\int d^{10} x {\> \rm Tr   }
\left\{
{1\over 4}Y_{mn}Y^{mn}
+{1\over 2}\bar \phi\left(\Gamma^m \partial_m \phi+\left[X_m,\, \phi\right]_- \right)\right\}, 
\label{action}
\eeq
\beq
Y_{mn}=\partial_mX_n-\partial_nX_m +\left[X_m,\, X_n\right]_-.
\label{F0def} 
\eeq
The notation is as follows. $X_m(\underline x)$ is the vector potential, $\phi(\underline x) $ is
the Majorana-Weyl spinor. Both are matrices in the adjoint representation of the gauge group 
${\bf G}$.  Latin indices
$m=0,\ldots 9$ describe Minkowski components.  Greek indices $\alpha=1,\ldots 16$ denote spinor
components. We will use the superspace formulation with odd coordinates $\theta^\alpha$. The  super
vector potentials, which are valued in the gauge group, are noted  
$A_m\left(\underline x,\underline \theta\right)$, $A_\alpha\left(\underline x,\underline
\theta\right)$. As discussed in ref.\cite{AFJ88}, we may
remove all the additional fields and uniquely reconstruct the physical fields $X_m$, $\phi$ from
$A_m$ and $A_\alpha$ if we impose the condition $\theta^\alpha A_\alpha=0$ on the latter.

With this condition, it was shown in refs\cite{W86}, \cite{AFJ88}, that the field equations derived
from the Lagrangian  \ref{action} are equivalent to the flatness conditions 
\beq
F_{\alpha \beta=0}, 
\label{flat}
\eeq
where $F$ is the supercurvature 
\beq
F_{\alpha \beta}=D_\alpha A_\beta+D_\beta A_\alpha+\left[A_\alpha,\, A_\beta\right]+
2\left(\sigma^m\right)_{\alpha\beta}A_m.  
\label{curdef}
\eeq
 $D_\alpha$ denote the superderivatives
\beq
D_\alpha=\partial_\alpha-\left(\sigma^m\right)_{\alpha \beta} 
\theta^\beta {\partial_m}, 
\label{sddef}
\eeq
and we use the Dirac matrices 
\beq
\Gamma^m=\left(\begin{array}{cc}
0_{16\times16}&\left(\left(\sigma^m\right)^{\alpha\beta}\right)\\
\left(\left(\sigma^m\right)_{\alpha\beta}\right)&0_{16\times16}
\end{array}\right),\quad  
\Gamma^{11}= \left(\begin{array}{cc}
1_{16\times16}&0\\0&-1_{16\times16}\end{array}\right).
\label{real1}
\eeq
The physical fields  appearing in equation \ref{action} are reconstructed from the superfields $A_m$
$A_\alpha$  as follows. Using the Bianchy identity on the super curvature one shows that 
one may write  
$$
F_{\alpha m}=\left(\sigma_m\right)_{\alpha \beta}\chi^{\beta}. 
$$ 
Then  $X_m$, $\phi^\alpha$ are, respectively,  the zeroth order contributions in the expansions of 
$A_m$ and $\chi^{\alpha}$ in powers of the odd coordinates $\theta$.

Throughout the paper, it will be convenient to use the following particular realisation: 
\beq
\left(\left(\sigma^{9}\right)^{\alpha\beta}\right)=
\left(\left(\sigma^{9}\right)_{\alpha\beta}\right)=
\left(\begin{array}{cc}
-1_{8\times 8}&0_{8\times 8}\\
0_{8\times 8}&1_{8\times 8}
\end{array}\right)
\label{real2}
\eeq
\beq
\left(\left(\sigma^{0}\right)^{\alpha\beta}\right)=-
\left(\left(\sigma^{0}\right)_{\alpha\beta}\right)=
\left(\begin{array}{cc}
1_{8\times 8}&0_{8\times 8}\\
0_{8\times 8}&1_{8\times 8}
\end{array}\right)
\label{real3}
\eeq
\beq
\left(\left(\sigma^{i}\right)^{\alpha\beta}\right)=-
\left(\left(\sigma^{i}\right)_{\alpha\beta}\right)=\left(\begin{array}{cc}
0&\gamma^i_{\mu,\overline \nu}\\
\left(\gamma^{i\, T}\right)_{\nu,\overline \mu}&0
\end{array}\right),\quad  i=1,\ldots 8. 
\label{real4}
\eeq
The convention for greek letters is as follows: Letters from the beginning of the alphabet run from
1 to 16. Letters from the middle of alphabet run from 1 to 8. In this way,  we shall separate
the two spinor representations of $O(8)$ by rewriting $\alpha_1,\ldots, \alpha_{16} $  as 
$\mu_1,\ldots, \mu_8, \overline \nu_1,\ldots, \overline \nu_8$
   
Using the above explicit realisations on sees that the equations to solve take the form  
\begin{eqnarray}
D_\mu A_\nu+D_\nu A_\mu +\left[A_\mu,\,
A_\nu\right]_+&=&2\delta_{\mu\nu}\left(A_0+A_9\right)\label{dynuu}\\
D_{\overline \mu} A_{\overline \nu}+D_{\overline \nu} A_{\overline \mu} +\left[A_{\overline \mu},\,
A_{\overline \nu}\right]_+&=&
2\delta_{{\overline \mu}{\overline \nu}}\left(A_0-A_9\right)\label{dyndd}\\  
D_{ \mu} A_{\overline \nu}+D_{\overline \nu} A_{ \mu} +\left[A_{\mu},\,
A_{\overline \nu}\right]_+&=&-2\sum_{i=1}^8 A_i\gamma^i_{\mu,\overline \nu}\label{dynud}
\end{eqnarray}
A lax pair formalism follows by noticing that 
if $\lambda^m\lambda_m=0$, there exists  
$R[\lambda]\in $ gauge group,  such that   
$$
\lambda^m \left(\sigma_m\right)^{\alpha \beta}
\left(A_\beta-R[\lambda]^{-1}D_\beta R[\lambda]\right)=0
$$
$$
\lambda^m \left(A_m-R[\lambda]^{-1}\partial_m R[\lambda]
\right)=0.  
$$
This allows us to express the solution in terms of pure gauges, but
this is not so useful 
since different
components are expressed in terms of $R[\lambda]$ involving different $\lambda$'s. 
 The drawback is that  $\lambda$,   which
plays the role of the spectral parameter, is a vector. In
particular, let us choose
$\lambda^{(\pm)}$, such that $\lambda^{(\pm)0}=\pm
\lambda^{(\pm)9}=1/2$, 
$\lambda^{(\pm)i}=0$, $i=1,\ldots 8$. This gives 
\beq
A_{ \mu}=R_+^{-1}D_\mu  R_+,\quad  A_+=R_+^{-1}\partial_+ R_+
\label{R+def}
\eeq
\beq
A_{ \overline \mu}=R_-^{-1}D_{\overline \mu } R_-,\quad  A_-=R_-^{-1}\partial_- R_-
\label{R-def}
\eeq
We let from now on 
\beq
A_{\pm}={A_0}\pm { A_9},\quad 
\partial_{\pm}={\partial\over \partial x^0}\pm {\partial\over \partial x^9}. 
\label{pmdef}
\eeq
One sees that $A_\mu$, $A_+$ (resp. $A_{\overline \mu}$, $A_-$) are expressed in terms of 
$R_+$ (resp. $R_-$). Moreover, since  only  the
$\mu \nu $ and $\overline\mu
\overline \nu$ are solved by the above there remain the  
 $\mu \overline \nu$ equations. A straightforward computation shows that they become 
$$
D_{\overline\nu}\left( {\bf R}^{-1}\, D_{\mu}\, {\bf R}\right) =-2\sum_{i=1}^8\tilde{A}_i
\gamma^i_{\mu,\overline\nu}
$$
$$
{\bf R}\equiv R_+R_-^{-1},\quad 
\tilde{A}_i\equiv R_-(A_i+\partial_i) R_-^{-1}
$$ 
We may derive  the field $\tilde A_i$, if the following conditions hold
\beq
\sum_{\mu \overline \nu  }D_{\overline\nu}\left( {\bf R}^{-1}\, D_{\mu}\, {\bf R}\right)
\gamma^{ijk}_{{\mu \overline \nu  }}=0,
\quad 
1\leq  i<j< k\leq 8.
\label{Dcond}
\eeq 
These are complicated  non linear $\sigma$ model type equations in superspace which so far could
not be handled. This is basically why these reasonings and the Lax representation just summarised did
not allow yet  to construct any explicit nontrivial physically meaningfull solution. Conditions 
\ref{Dcond} only provide  a procedure\cite{AFJ88} for obtaining infinite series of nonlocal, and
rather complicated conservation laws. 
\section{A usefull on-shell gauge}
Under gauge transformations, we have  
$$
R_\pm \to R_\pm \Lambda, \quad 
A_m\to \Lambda^{-1}\left(A_m+\partial_m\right) \Lambda,\quad 
A_\alpha\to \Lambda^{-1}\left(A_\alpha+D_\alpha\right) \Lambda
$$
Thus ${\bf R}$ is gauge invariant. If $\Lambda=R_-^{-1}$, we get 
\beq
A_+\to{\bf R}^{-1}\partial_+{\bf R} ,\quad  A_\mu \to 
{\bf R}^{-1}D_\mu{\bf R},\quad 
 A_i\to \widetilde A_i
\label{osgauge}
\eeq   
$$
A_-\to 0, \quad A_{\overline \mu }\to 0. 
$$
Thus, if the field equations  are satisfied there exists a gauge (on shell) such that  
$A_-=A_{\overline \mu }=0$. After this gauge choice, the flatness conditions
\ref{dynuu}--\ref{dynud} boil down to
\begin{eqnarray}
D_\mu A_\nu+D_\nu A_\mu +\left[A_\mu,\,
A_\nu\right]_+&=&4\delta_{\mu\nu}A_0\label{dynuu0}\\
0&=&0\label{dyndd0}\\
D_{\overline \nu} A_{ \mu} &=&-2\sum_1^8 \widetilde A_i\gamma^i_{\mu,\overline \nu}
\label{dynud0}
\end{eqnarray}
The last  mixed ones which in general lead to the complicated conditions \ref{Dcond} have become
linear, and we will next derive their general solution.  In order to do so, we will use the
following explicit realisation of the $O(8)$ Dirac matrices, 
\begin{eqnarray}
\gamma^1= \tau_1\otimes \tau_3\tau_1\otimes {\bf 1} \quad &\quad 
\gamma^5=\tau_3\otimes \tau_3\tau_1\otimes {\bf 1} \nonumber\\
\gamma^2= {\bf 1}\otimes \tau_1\otimes \tau_3\tau_1 \quad &\quad
\gamma^6= {\bf 1}\otimes \tau_3\otimes \tau_3\tau_1   \nonumber\\
\gamma^3=\tau_3\tau_1 \otimes {\bf 1}\otimes \tau_1 \quad &\quad
\gamma^7= \tau_3\tau_1\otimes {\bf 1} \otimes \tau_3  \nonumber\\
\gamma^4= \tau_3\tau_1\otimes \tau_3\tau_1\otimes \tau_3\tau_1 \quad &\quad
\gamma^8={\bf 1}\otimes  {\bf 1}\otimes {\bf 1}. 
\label{gamdef}
\end{eqnarray}
Substituting into Eq.\ref{dynud0} we get 
\begin{eqnarray}
-2 \widetilde A_1&=f_{\overline 17}=f_{\overline 28}=f_{\overline 53}=f_{\overline 64} \nonumber\\
-2 \widetilde A_2&=f_{\overline 14}=f_{\overline 32}=f_{\overline 58}=f_{\overline 76}\nonumber\\
-2 \widetilde A_3&=f_{\overline 16}=f_{\overline 25}=f_{\overline 38}=f_{\overline 47}\nonumber\\
-2 \widetilde A_4&=f_{\overline 18}=f_{\overline 45}=f_{\overline 63}=f_{\overline 72}\nonumber\\
-2 \widetilde A_5&=f_{\overline 13}=f_{\overline 24}=f_{\overline 75}=f_{\overline 86}\nonumber\\
-2 \widetilde A_6&=f_{\overline 12}=f_{\overline 43}=f_{\overline 56}=f_{\overline 87}\nonumber\\
-2 \widetilde A_7&=f_{\overline 15}=f_{\overline 62}=f_{\overline 37}=f_{\overline 84}
\label{mix1}  
\end{eqnarray}
\beq
-2\widetilde A_8=f_{\overline \mu \mu},\quad \mu=1,\ldots 8
\label{mix2}
\eeq
\beq
f_{\overline \mu \nu}=-f_{\overline\nu \mu}
\label{mix3}
\eeq
where we have let $f_{\overline \mu \nu}=D_{\overline \mu} A_\nu$
By convention overlined and non overlined indices with the same letter (such as $\mu$ and
$\overline \mu$) take the same numerical value. With the particular realisation of 
$\sigma$ matrices displayed on Eqs.\ref{real1}--\ref{real4}, one has the anticommutation relations
\beq
\left[D_\mu,\, D_\nu\right]_+=2 \delta_{\mu \nu}\partial_+,\quad 
\left[D_{\overline \mu},\, D_{\overline \nu}\right]_+=
2 \delta_{{\overline \mu} {\overline \nu}}\partial_-
\label{ant1}
\eeq
Thus it  follows from   equation \ref{mix3} that 
 there exits a superfield  $\Phi$ such that 
\beq
A_\nu=D_{\overline \nu }\Phi
\label{Phidef}
\eeq
 Then equation \ref{mix2} is automatically satisfied since it becomes  $f_{\overline \mu
\mu}=D^2_{\overline\mu}\Phi=\partial_- \Phi$ which is indeed  independent  from
$\overline \mu$.
 On each line, the corresponding
component of the vector potentail $\widetilde A_i$ may be computed iff the three right most
equalities are satisfied. Thus we have the consistency equations of the superfield $\Phi$

\begin{eqnarray}
D_{\overline 1}D_{\overline  7}\Phi=  D_{\overline 2}D_{\overline  8}\Phi=  
D_{\overline 5}D_{\overline  3}\Phi=  D_{\overline 6}D_{\overline  4}\Phi\nonumber \\ 
D_{\overline 1}D_{\overline  4}\Phi=  D_{\overline 3}D_{\overline  2}\Phi=  
D_{\overline 5}D_{\overline  8}\Phi=  D_{\overline 7}D_{\overline  6}\Phi\nonumber\\
D_{\overline 1}D_{\overline  6}\Phi=  D_{\overline 2}D_{\overline  5}\Phi=  
D_{\overline 3}D_{\overline  8}\Phi=  D_{\overline 4}D_{\overline  7}\Phi\nonumber\\
D_{\overline 1}D_{\overline  8}\Phi=  D_{\overline 4}D_{\overline  5}\Phi=  
D_{\overline 6}D_{\overline  3}\Phi=  D_{\overline 7}D_{\overline  2}\Phi\nonumber\\
D_{\overline 1}D_{\overline  3}\Phi=  D_{\overline 2}D_{\overline  4}\Phi=  
D_{\overline 7}D_{\overline  5}\Phi=  D_{\overline 8}D_{\overline  6}\Phi\nonumber \\
D_{\overline 1}D_{\overline  2}\Phi=  D_{\overline 4}D_{\overline  3}\Phi=  
D_{\overline 5}D_{\overline  6}\Phi=  D_{\overline 8}D_{\overline  7}\Phi\nonumber \\
D_{\overline 1}D_{\overline  5}\Phi=  D_{\overline 6}D_{\overline  2}\Phi=  
D_{\overline 3}D_{\overline  7}\Phi=  D_{\overline 8}D_{\overline  4}\Phi
\label{self8}  
\end{eqnarray}
These consistency conditions take the form 
$$
D_{\overline \mu}D_{ \overline \nu}\Phi=\sum_{\overline \rho< \overline\lambda}T_{\overline \mu
\overline \nu \overline \rho \overline
\lambda}D_{\overline \rho}D_{ \overline
\lambda}\Phi. 
$$
where $T_{\overline \mu \overline \nu \overline \rho \overline
\lambda}$ is a numerical tensor which is antisymmetric with elements equal to  $\pm 1$, or
$0$. Thus we have equations similar to the self--duality relations considered, for bosonic
variables in ref.\cite{CDFN83}.   Here the main difference is that we have
superderivatives and that our equations are linear partial differential equations. 
\section{Solution of the reduced self--duality equations}  
Let us derive the general solution of equations \ref{mix1} in the particular reduced case where 
$\Phi $ does not depend upon $x^i$, for $i=1,\ldots 8$. Then the superderivatives take the form 
\beq
D_{\mu}={\partial\over \partial \theta^{\mu}}+\theta^{\mu}\partial_+,\quad 
D_{\overline \mu}={\partial\over \partial \theta^{\overline \mu}}+\theta^{\overline \mu}\partial_-
\label{redsd}
\eeq 
In this case equations \ref{self8}  only involve the variables $x_-, \theta^{\overline 1}\ldots 
\theta^{\overline 8} $,
and we forget the other variables for the time being. 
 In the forthcoming discussion,  we will find it useful to
use the following lemna
\subsection{Super Cauchy relations:} Consider any pair of different indices, which are selected once
for all in this subsection, say $\overline \mu $, $\overline \nu$.  Given an arbitary superfield
$\Upsilon(x_-, \theta^{\overline 1}, 
\ldots \theta^{\overline 8})$, there exists a
superfield $\Lambda(x_-, \theta^{\overline 1}, 
\ldots \theta^{\overline 8})$ such that 
\beq
D_{\overline \mu }\Upsilon =D_{\overline \nu}\Lambda,  \qquad D_{\overline \nu }\Upsilon
=-D_{\overline \mu}\Lambda  
\label{scauchy}
\eeq
\paragraph{Proof:}
First one verifies that the consistency of these equations is a consequence of the
equations $D^2_{\overline \mu }=D^2_{\overline \nu }$, and  
$\left[D_{\overline \mu },\, D_{\overline \nu }\right]_+=0$ which follow from the superalgebra.
Next, one may explicitly solve order by order in the expansion in powers of $\theta^{\overline
\mu}$. For an arbitrarily given superfield with the expansion
\beq
F(x_-, \theta^{\overline 1},\ldots, \theta^{\overline 8})= \sum_{p=0}^8\sum_{\overline \mu_1,\ldots,
\overline \mu_p} {\theta^{\overline  \mu_1}\cdots
\theta^{\overline  \mu_p}\over p !}F^{(p)}_{\overline \mu_1\ldots \overline \mu_p}(x_-), 
\label{exp}
\eeq  
the superderivatives act as follows, (with $p=1,\ldots, 7$), 
\beq
\left(D_{\overline \mu} F\right)^{(p)}_{\overline \mu_1,\ldots, \overline \mu_p}
=F^{(p+1)}_{\overline \mu\, \overline \mu_1\ldots \overline \mu_p}+
\sum_{i=1}^p\left(-1\right)^{i+1}\delta_{\overline \mu, \overline \mu_i}
\partial_- F^{(p-1)}_{\overline \mu_1\ldots /\!\!\!\! \overline \mu_i \ldots  \overline
\mu_p}
\label{Ddef2}
\eeq
where the extremal values of $p$ are treated by setting $F^{(p)}=0$ if $p<0$ or if $p>8$. 
This allows us to solve order by order. One finds the relations 
$$
\Upsilon^{(p)}_{\overline \mu\, \overline \mu_1\ldots \overline \mu_{p-1}}=
\Lambda^{(p)}_{\overline \nu\, \overline \mu_1\ldots \overline \mu_{p-1}},\quad 
\Upsilon^{(p)}_{\overline \nu\, \overline \mu_1\ldots \overline \mu_{p-1}}=
-\Lambda^{(p)}_{\overline \mu\, \overline \mu_1\ldots \overline \mu_{p-1}},\quad 
p=1,\ldots, 8
$$
\beq
\partial_- \Upsilon^{(p)}_{\overline \mu_1\ldots  \overline
\mu_p}=-
\Lambda^{(p+2)}_{\overline \mu \overline \nu\, \overline \mu_1 \ldots
\overline \mu_p},
\quad  
\Upsilon^{(p+2)}_{\overline \mu \overline \nu\, \overline \mu_1 
\ldots\overline \mu_p}= 
\partial_- \Lambda^{(p)}_{\overline \mu_1\ldots   \overline
\mu_p},p=0\ldots,6, 
\label{upcauch}
\eeq
which  determine $\Lambda$ once $\Upsilon$ is given. 
 \subsection{Selfduality in four variables }
In order to derive the general solution of equations \ref{self8},   we remark that we may
re-arrange these relations under the form  
\beq
\begin{array}{llll}
D_{\overline   1\overline  7}\Phi&=D_{\overline   2\overline  8}\Phi,\quad  
D_{\overline   1\overline  2}\Phi&=D_{\overline   8\overline  7}\Phi, \quad 
D_{\overline   1\overline  8}\Phi&=D_{\overline   7\overline  2}\Phi;  \\
D_{\overline   1\overline  7}\Phi&=D_{\overline   5\overline  3}\Phi,\quad  
D_{\overline   1\overline  5}\Phi&=D_{\overline   3\overline  7}\Phi, \quad 
D_{\overline   1\overline  3}\Phi&=D_{\overline   7\overline  5}\Phi;    \\
D_{\overline   1\overline  7}\Phi&=D_{\overline   6\overline  4}\Phi,\quad  
D_{\overline   1\overline  6}\Phi&=D_{\overline   4\overline  7}\Phi, \quad 
D_{\overline   1\overline  4}\Phi&=D_{\overline   7\overline  6}\Phi; \\
D_{\overline   1\overline  4}\Phi&=D_{\overline   3\overline  2}\Phi,\quad  
D_{\overline   1\overline  3}\Phi&=D_{\overline   2\overline  4}\Phi, \quad 
D_{\overline   1\overline  2}\Phi&=D_{\overline   4\overline  3}\Phi;\\
D_{\overline   1\overline  4}\Phi&=D_{\overline   5\overline  8}\Phi,\quad  
D_{\overline   1\overline  5}\Phi&=D_{\overline   8\overline  4}\Phi, \quad 
D_{\overline   1\overline  8}\Phi&=D_{\overline   4\overline  5}\Phi; \\
D_{\overline   1\overline  6}\Phi&=D_{\overline   2\overline  5}\Phi,\quad  
D_{\overline   1\overline  2}\Phi&=D_{\overline   5\overline  6}\Phi, \quad 
D_{\overline   1\overline  5}\Phi&=D_{\overline   6\overline  2}\Phi; \\
D_{\overline   1\overline  6}\Phi&=D_{\overline   3\overline  8}\Phi,\quad  
D_{\overline   1\overline  3}\Phi&=D_{\overline   8\overline  6}\Phi, 
\quad D_{\overline  1\overline  8}\Phi&=D_{\overline   6\overline  3}\Phi. 
\end{array}
\label{self84}
\eeq
From now on we let $D_{\overline \mu \overline \nu}=D_{\overline \mu}D_{ \overline \nu}$. 
Each line  forms a closed set of  self duality equations in four
variables.  Thus we first solve this type of equations.
Consider, for fixed 
$\overline \mu\not=\overline\nu \not=\overline \sigma\not=\overline\rho$, the equations
\beq
\left(D_{\overline \mu\overline \rho}-D_{\overline \sigma\overline \nu} \right)\Phi=0,\quad 
\left(D_{\overline \nu\overline \rho}-D_{\overline \mu\overline \sigma}\right)\Phi=0,\quad 
\left(D_{\overline \mu\overline \nu}-D_{\overline \rho\overline \sigma}\right)\Phi=0.  
\label{self4}
\eeq
By using the superalgebra \ref{ant1}, one finds that for arbitrary superfield  
$\Psi_1$, $\Phi=\left(D_{\overline \mu}D_{\overline \rho}+D_{\overline \sigma}D_{\overline \nu}
\right)\Psi_1$ is a solution of these last three equations. This form is suspiciously non
symmetric.  However, we use the above super Cauchy relations to introduce superfields $\Psi_2$,
$\Psi_3$  such that 
\begin{eqnarray}
D_{\overline \mu }\Psi_1 =D_{\overline \nu}\Psi_2 &, &D_{\overline \nu }\Psi_1
=-D_{\overline \mu}\Psi_2 \nonumber\\
D_{\overline \rho }\Psi_1 =D_{\overline \nu}\Psi_3 &,&D_{\overline \nu }\Psi_1
=-D_{\overline \rho}\Psi_3\nonumber\\
\label{sc2}
\end{eqnarray}
Then we see that we may write our solution under three equivalent forms 
\beq
\Phi=\left(D_{\overline \mu}D_{\overline \rho}+D_{\overline \sigma}D_{\overline \nu}
\right)\Psi_1 
=\left(D_{\overline \nu}D_{\overline \rho}+D_{\overline \mu}D_{\overline \sigma}\right)\Psi_2
=\left(D_{\overline \mu}D_{\overline \nu}+D_{\overline \rho}D_{\overline \sigma}\right)\Psi_3.
\label{sol4}
\eeq 
where the expected symmetry becomes manifest.  



\subsection{Eight variables, the Cartan basis:}
Actually, the superalgebra satisfied by the $D_{\overline \mu}$ operators, as displayed by
equations \ref{ant1} coincides with a Dirac algebra in eight dimensions up to the $\partial_-$ 
differential operator. It thus
follows that the super derivatives $D_{\overline \mu \overline \nu}$ obey an $so(8)$ Lie
algebra\footnote{Since the $\partial_-$ factor will not play a significant role in the forthcoming
argument, we do not speak about  it any longer  for brevity.}. It is useful to organise the
superderivatives appearing in the self--duality relations in a  Cartan basis. For this we
temporarily re-label the indices as follows: $\left(\overline 1,\ldots,\overline 8\right)\to
\left(\overline 1,-\overline 1,\overline 2,-\overline 2,\overline 3,-\overline 3,\overline
4,-\overline 4\right)$. The roots of $so(8)$ may be written under the form 
$\pm \vec e_i\pm \vec e_j$, $1\leq j<k\leq 4$. Let us denote $E_{\pm \vec e_i\pm \vec e_j}$ the 
step operators and by $h_{\pm \vec e_i\pm \vec e_j}$ the Cartan generators. One finds the
correspondence
\begin{eqnarray}
E_{\vec e_j\pm \vec e_k}+E_{-\vec e_j \mp \vec e_k}&=&
{\mp i\over 2}\left( D_{-\overline \jmath  \,\overline k }\pm  D_{\overline  \jmath\, -\overline  k
}\right)\nonumber\\ E_{\vec e_j\pm  \vec e_k}-E_{-\vec e_j \mp \vec e_k}&=&
{\mp 1\over 2} \left(D_{\overline \jmath \overline k} \mp  D_{-\overline \jmath\, -\overline
k}\right)\nonumber\\ h_{\vec e_j\pm  \vec e_k}&=&
{i\over 2}\left(D_{\overline \jmath\, -\overline \jmath}\pm D_{\overline  k\,  -\overline
k}\right)\label{cartan}
\end{eqnarray}
with the convention that numerically $j=\overline \jmath$, $k=\overline k$. 

Out of the seven self dualities in four variables, six may be written as 
\beq
D_{\overline \jmath \, -\overline k}\Phi =D_{\overline k\, -\overline \jmath }\Phi,\quad  
D_{\overline
\jmath\,  
\overline k}\Phi=D_{-\overline \jmath\,  -\overline k}\Phi, \quad 
D_{\overline \jmath \, -\overline \jmath }\Phi=D_{-\overline k\, \overline k}\Phi
\label{cartan2}
\eeq
with $1\leq \overline \jmath < \overline k\leq 4$. The corresponding differential operators 
generate an
$su(4)$ algebra with simple roots $\vec e_1+\vec e_2$, $\vec e_2+\vec e_3$, $\vec e_3+\vec e_4$.
Out of these six triplets of relations only the three associated with the simple roots are
independent. 
In total,  we are thus left with four triplets of independent relations        

\subsection{Solving the $su(4)$ part}
Consider a particular simple root $\vec e_i+\vec e_{i+1}$. The corresponding self--duality relations
take the form 
$$
E_{\vec e_i+\vec e_{i+1}}\Phi=E_{-\vec e_i-\vec e_{i+1}}\Phi=h_{\vec e_i+\vec e_{i+1}}\Phi=0. 
$$
and the general solution derived above becomes
$$
\Phi= E_{\vec e_i-\vec e_{i+1}}\Psi_1=E_{-\vec e_i+\vec e_{i+1}}\Psi_2=h_{\vec e_i-\vec
e_{i+1}}\Psi_3. 
$$
Group theoretically, the super Cauchy relations are seen to allow us to use any of the three
generators of the $su(2)$ subalgebra with root $\vec e_i-\vec e_{i+1}$. Returning to the
full $su(4)$  part, we use this liberty to pick up the Cartan generator for each simple root.
This leads us to the ansatz  
\beq
\Phi=h_{e_1-e_2}h_{e_2-e_3}h_{e_3-e_4}\widetilde \Phi. 
\label{su4ans}
\eeq
Since the $h$'s commute, 
it should be clear from the above that, for arbitrary $\widetilde \Phi$,   the last expression 
obeys all six self--duality relations associated with the $su(4)$ mentioned above. This may of
course be checked explicitly from the superalgebra. Let us return to the previous label of indices.
At this point it is convenient to introduce the equivalent of the nineth Dirac matrix by writing  
\beq
D_{\overline 9}=D_{\overline 1}D_{\overline 2}\ldots D_{\overline 8},
\label{d9def}
\eeq 
which is  such that 
$D^2_{\overline 9}=\partial^8_-$. After some calculations one may show that above may be written
as 
\beq
\Phi=\left\{D_{\overline 1\overline 2}-D_{\overline 3 \overline  4} 
+ D_{\overline 5\overline 6} - D_{\overline 7\overline 8}\right\}\chi , 
\label{ans2}
\eeq
where we have let 
\beq
\chi=\left(D_{\overline 9}+ \partial^4_-\right)\widetilde \Phi. 
\label{ans3}
\eeq
This exhibits a chirality projector which commutes with all the $D_{\overline \mu \overline
\nu}$. 
 \subsection{Solving the last set of self-duality relations }
 Looking at the last set of relations, one sees that our task is to determine $\chi$ such that  
\begin{eqnarray}
\left(D_{{\overline 2}{\overline 8}}-D_{{\overline 5}{\overline 3}}\right)
\left(D_{{\overline 1} {\overline 2}}-D_{{\overline 3}{\overline 4}} 
+ D_{{\overline 5}{\overline 6}} - D_{{\overline 7}{\overline 8}}\right)\chi
&=&0
\label{leq1}\\
\left(D_{{\overline 3}{\overline 2}} -D_{{\overline 5}{\overline 8}}\right)
\left(D_{{\overline 1}{\overline 2}}-D_{{\overline 3}{\overline 4}} 
+ D_{{\overline 5}{\overline 6}} -D_{{\overline 7}{\overline 8}}\right)\chi 
&=&0
\label{leq2}\\ 
\left(D_{{\overline 2}{\overline 5}}-D_{{\overline 3}{\overline 8}}\right)
\left(D_{{\overline 1} {\overline 2}}-D_{{\overline 3}{\overline 4}} 
+ D_{{\overline 5}{\overline 6}} - D_{{\overline 7}{\overline 8}}\right)\chi
&=&0
\label{leq3}
\end{eqnarray}
Let us first solve the first equation separately. We will see that at the end the other two will
also be satisfied. We will use  again  super Cauchy relations of the type \ref{scauchy}. One easily
verifies that one may at the same time apply two super Cauchy  transformations over independent
variables. It is convenient to introduce a superfield $\chi_{1}$ such that
\beq
D_{\overline 2}  \chi=D_{\overline 7}  \chi_1,\quad 
D_{\overline 7}  \chi=-D_{\overline 2}  \chi_1, 
\label{cauch1}
\eeq
\beq
D_{\overline 4}  \chi=D_{\overline 5}  \chi_1,\quad 
D_{\overline 5}  \chi=-D_{\overline 4}  \chi_1, 
\label{cauch2}
\eeq 
we get  
$$
\left(D_{{\overline 2}{\overline 8}}-D_{{\overline 5}{\overline
3}}\right)
\left(D_{{\overline 1} {\overline 7}} + D_{{\overline 4} {\overline 6}} \right)
\chi_1   =0. 
$$
Thus the general solution is   
$$
 \chi_1=\left(D_{{\overline 2}{\overline 8}}+D_{{\overline 5}{\overline 3}}\right) F_1+
\left(D_{{\overline 4} {\overline 6}} -  D_{{\overline 1} {\overline 7}}\right) G_1,  
$$  
where $F_1$ and $G_1$ are arbitrary chiral superfields.  
Is it possible to go back to $ \chi$ ?
We apply Cauchy equations to each term, by letting  
\beq
D_{\overline 2}  F=D_{\overline 7}  F_1,\quad 
D_{\overline 7}  F=-D_{\overline 2}  F_1; \qquad 
D_{\overline 2}  G=D_{\overline 7}  G_1,\quad 
D_{\overline 7}  G=-D_{\overline 2}  G_1; 
\label{cauch3}
\eeq
\beq
D_{\overline 4}  F=D_{\overline 5}  F_1,\quad 
D_{\overline 5}  F=-D_{\overline 4}  F_1;\qquad 
D_{\overline 4}  G=D_{\overline 5}  G_1,\quad 
D_{\overline 5}  G=-D_{\overline 4}  G_1. 
\label{cauch4}
\eeq 
Then one may verify that equations \ref{cauch1} \ref{cauch2} are satisfied with
\beq
  \chi=\left(-D_{{\overline 2}{\overline 8}}+D_{{\overline 5}{\overline 3}}\right) F+
\left(D_{{\overline 1} {\overline 7}} +D_{{\overline 4}, {\overline 6}} \right) G. 
\label{ans01}
\eeq 
At this point, there appears an great simplification since one may check that 
$$
\left\{D_{{\overline 1}{\overline 2}}-D_{{\overline 3} {\overline 4}} 
+ D_{{\overline 5} {\overline 6}} - D_{{\overline 7}{\overline 8}}\right\}\left(-D_{{\overline
2}{\overline 8}}+D_{{\overline 5}{\overline 3}}\right)
$$
\beq
=\left\{D_{{\overline 1}{\overline 2}}-D_{{\overline 3} {\overline 4}} + D_{{\overline 5} {\overline
6}} - D_{{\overline 7}{\overline 8}}\right\}\left(D_{{\overline 1} {\overline 7}} +D_{{\overline
4} {\overline 6}} \right). 
\label{simpl}
\eeq
Thus we may forget the $G$ term. Turning finally to equations \ref{leq2}, and \ref{leq3}
one may explicitly verify that they are  automatically satisfied for arbitrary $F$ since one has 
\beq
\left(D_{{\overline 3}{\overline 2}}-D_{{\overline 5}{\overline
8}}\right)
\left\{D_{{\overline 1}{\overline 2}}-D_{{\overline 3}
{\overline 4}} + D_{{\overline 5} 
{\overline 6}} - D_{{\overline 7}{\overline 8}}\right\}
\left(-D_{{\overline 2}{\overline 8}}+D_{{\overline 5}{\overline 3}}\right)=0, 
\label{simpl2}
\eeq
\beq
\left(D_{{\overline 2}{\overline 5}}-D_{{\overline 3}{\overline
8}}\right)
\left\{D_{{\overline 1} {\overline 2}}-D_{{\overline 3} {\overline 4}}
+ 
D_{{\overline 5} {\overline 6}} - D_{{\overline 7}{\overline 8}}\right\}
\left(-D_{{\overline 2}{\overline 8}}+D_{{\overline 5}{\overline 3}}\right)=0. 
\label{simpl3}
\eeq
Altogther, we have shown that the general solution of the eight self--duality relations 
\ref{self8}
is given by 
\beq
\Phi=\left\{D_{{\overline 1} {\overline 2}}-D_{{\overline 3}
{\overline 4}} 
+ D_{{\overline 5} {\overline 6}} - D_{{\overline 7}{\overline 8}}\right\}
\left(D_{{\overline 5}{\overline 3}}-D_{{\overline 2}{\overline 8}}\right)\left(D_{\overline 9}
+ \partial^{4}_-\right) \Psi,  
\label{solself8}
\eeq
where we have let 
$F =\left(D_{\overline 9}+ \partial^{4}_-\right)\Psi$ in agreement with equation 
\ref{ans3}, and $\Psi$ is an arbitrary superfield. At this point, we have to admit that we are
unable to explain why the key relations
\ref{simpl}--\ref{simpl3} hold, apart from verifying them explicitly. They where discovered using
Mathematica on the analogous relations for $O(8)$ Dirac matrices. The form of the solution is not
explicitly symmetric between indices, but here also, as for the above case of four variables,
symmetry may be verified from super Cauchy transformations.      
\section{The non linear equations for $\Phi$.}
 
Equations \ref{dynuu0} remain to be solved. 
Make use of equation \ref{Phidef} and substitute\footnote{As before, numerically $\mu=\overline
\mu$, 
$\nu=\overline \nu$.} 
  $A_\mu =D_{\overline \mu} \Phi$, $A_\nu =D_{\overline \nu} \Phi$.
One gets 
\beq
D_\mu D_{\overline \nu}\Phi+D_\nu D_{\overline \mu}\Phi + 
\left[D_{\overline \mu}\Phi,\,  D_{\overline
\nu}\Phi\right]_+=4\delta_{\mu \nu} A_0. 
\label{nonleq}
\eeq
The super field $A_0$ may be computed from these equations if the following consistency conditions
hold.   We must verify that, for $\mu\not=\nu$,  
\beq
D_\mu D_{\overline \nu}\Phi+D_\nu D_{\overline \mu}\Phi + 
\left[D_{\overline \mu}\Phi,\, D_{\overline
\nu}\Phi\right]_+=0,  
\label{syang}
\eeq
and that 
\beq
D_\mu D_{\overline \mu}\Phi+ \left(D_{\overline \mu}\Phi\right)^2 
{\> \rm is\> independent\> from   \> } \mu. 
\label{syang2}
\eeq
If these conditions hold, 
the superfield $A_0$ may be computed from   
\beq
A_0={1\over 2}\left\{D_1 D_{\overline 1}\Phi+ \left(D_{\overline 1}\Phi\right)^2\right\}. 
\label{azdef}
\eeq
At this point it is interesting to recall the  four dimensional Yang equations which arose in
solving self--dual (purely bosonic) Yang--Mills in four dimensions. For this, we closely
follow the review\cite{LS89}.  There are two bosonic complex coordinates $z$, $y$ and their
conjugate
$\bar z$, $\bar y$. One may start from the equations  (Indices mean derivatives)
$$
\left(G_zG^{-1}\right)_{\bar z}+\left(G_yG^{-1}\right)_{\bar y}=0,
$$
where $G$  is in the adjoint representation of the gauge group. This is partially solved by letting 
\beq
G_{ z}G^{-1}=f_{\bar y},\quad G_{ y}G^{-1}=-f_{\bar z}.
\label{fdef}
\eeq
which leads to the consistency condition 
\beq
f_{\bar z z}+f_{\bar y y}+\left[f_{\bar y},\, f_{\bar z}\right]_-=0.
\label{feq}
\eeq
In order to draw a parallel with our case, let us recall that, according to equations 
\ref{osgauge}, \ref{Phidef}, we have 
\beq
A_\mu={\bf R}^{-1}D_\mu {\bf R}=D_{\overline \mu} \Phi. 
\label{R-Phi}
\eeq
There is a similarity between equations \ref{nonleq} and \ref{feq}, and between
equations \ref{fdef}, and \ref{R-Phi}, except that the indices are paired differently. 
As we will next show, this fact, together with the basic properties of  superderivatives {
\bfseries allows us
to  discuss the solution of the present equations for any number of supervariables, while the bosonic
case may be handled only with four coordinates.  }
\subsection{A solution of the non linear consistency conditions}
In this part we adapt to our problem  the perturbative method, valid to all orders in the coupling
constant $g$, developed earlier for the bosonic case (see ref.\cite{LS89} and refs. therein).
We expand in powers of $g$ after replacing equation \ref{nonleq} by
\beq
D_\mu D_{\overline \nu}\Phi+D_\nu D_{\overline \mu}\Phi + 
g \left[D_{ \overline \mu}\Phi,\, D_{ \overline \nu}\Phi\right]_+=4\delta_{\mu \nu} A_0. 
\label{nonleqg}
\eeq  
Following a path very similar to the bosonic case (see ref.\cite{LS89}), one derives the following 
solution to all orders in $g$. Assume there exists a superfield 
$F\left(\lambda, x_+, x_-,\theta^1,\ldots, \theta^8, \overline \theta^1,\ldots,
\overline\theta^8\right)$, (noted $F(\lambda)$ for brevity) with $\lambda$ an arbitrary (bosonic)
parameter, such that 
\beq
D_{\mu}F(\lambda)=\lambda D_{\overline \mu}F(\lambda). 
\label{vraie}
\eeq 
We shall solve this equation later on explicitly. Then the solution may be written as 
\beq
\Phi=-\sum_{n=0}^\infty{ g^n\over (n+1)! } \int d\lambda F^{\left[n\right]}\left(\lambda\right), 
\label{pertdef}
\eeq
where $F^{\left[n\right]}$ is defined by the recursion 
\beq
F^{\left[n\right]}(\lambda)=
\sum_{p=0}^{n-1}{n-1\choose p} \left[F^{\left[p\right]}(\lambda),\, 
\int d\lambda' {F^{\left[n-1-p\right]}(\lambda')\over \lambda-\lambda'}\right]_-. 
\label{lsans}
\eeq 
with $F^{\left[0\right]}(\lambda)=F(\lambda)$. 
\paragraph{Proof:} The best method is to  first derive that the above
expression satisfies a first order differential equation of the form 
\beq
D_{\mu }\Phi=D_{\overline \mu}\Omega +g \Phi D_{\overline \mu}\Phi,  
\label{foeq}
\eeq
where $\Omega$ is a superfield which is computed order by order in $g$. Then it is easy to verify
that  conditions \ref{syang} and \ref{syang2} follow.  Checking equation \ref{foeq} is
straightforward but lengthy. The calculation is almost the same as in the bosonic case. Thus we
omit it.   Then it is easily verified that equations
\ref{azdef} hold with 
\beq
A_0={1\over 2}\left(\partial_-\Omega+g\Phi\partial_-\Phi\right)
\label{a0res}
\eeq
There remains to verify equation \ref{vraie}. First, applying $D_\mu$
to both sides of 
 equation \ref{vraie} 
one derives  the consistency condition 
\beq
\left(\partial_++\lambda^2\partial_- \right)F(\lambda)=0
\label{vrfx}
\eeq
  Next, start from the general expansion. 
$$
F(\lambda, x_+, x_-,\theta^{ 1},\ldots, \theta^{ 8},  \theta^{\overline 1},\ldots,
\theta^{\overline 8})= \sum_{p=0}^8\sum_{q=0}^8 
\sum_{\mu_1,\ldots,  \mu_p}\sum_{ \overline \nu_1,\ldots, \overline\nu_q}
$$
 \beq
{ \theta^{ \mu_1}\cdots \theta^{\mu_p} \theta^{\overline \nu_1}\cdots \theta^{\overline \nu_q}
\over p!  q !}  
F^{(p,q)}_{ \mu_1\ldots \mu_p\overline\nu_1\ldots \overline
\mu_q}(\lambda, x_+, x_-). 
\label{Fexpdef}
\eeq
Equations \ref{vraie} give 
$$
F^{(p+1,q)}_{\mu\, \mu_1\ldots \mu_p,\overline\nu_1,\ldots, \overline\nu_q}+
 \sum_{i=1}^p\left(-1\right)^{i+1}\delta_{\mu, \mu_i}
\partial_+ F^{(p-1,q)}_{\mu_1\ldots /\!\!\!\! \mu_i \ldots  
\mu_p,\overline\nu_1,\ldots, \overline\nu_q}
$$
\beq
=\lambda  \left(-1\right)^{p}
F^{(p,q+1)}_{\mu_1\ldots \mu_p,\overline \mu\, \overline\nu_1,\ldots, \overline\nu_q}
+\lambda \left(-1\right)^{p}
\sum_{i=1}^q\left(-1\right)^{i+1}\delta_{\overline \mu, \overline\nu_i}
\partial_- F^{(p,q-1)}_{\mu_1,\ldots  
\mu_p,\overline\nu_1, \ldots /\!\!\!\! \overline\nu_i ,\ldots, \overline\nu_q}. 
\label{vraie2}
\eeq
The general solution is as follows. Take all indices different, unless they are noted with the same
letters with and without overline. Then one has 
\beq
F^{(p+k,q+k)}_{\rho_1 \ldots \rho_k\, \mu_1\ldots \mu_p,\overline \rho_1 \ldots \overline  \rho_k
\overline\nu_1,\ldots,
\overline\nu_q}=
\lambda^{p+k} \left(-1\right)^{p(p+k) +k(k-1)/2} \partial^k_- 
F^{(0,q+p)}_{\overline \mu_1\,\ldots  \overline \mu_p \overline\nu_1,\ldots,
\overline\nu_q}
\label{vrfk}
\eeq
This, together, with the antisymmetry, allows to express all tensors in terms of 
$F^{(0,n)}_{\overline \mu_1\,\ldots  \overline \mu_n}$ which may be arbitrarily chosen as
functions of a single variable. 

For later purpose, we derive a compact expression of the solution. It is straightforward to verify
that, if we define 
\beq
\Theta(\lambda)=e^{\lambda \sum_\mu
\left(\theta_\mu D_{\overline
\mu}\right)},  
\label{Thetdef}
\eeq
we have
\beq
\Theta(\lambda) {\partial\over \partial \theta^{\mu}}\Theta^{-1}(\lambda)=
D_\mu-\lambda D_{\overline \mu}-\theta^{\mu}\left(\partial_++\lambda^2\partial_-\right)
\label{transf1}
\eeq
that we will satisfy the equation $D_\mu F(\lambda)=\lambda D_{\overline \mu}F(\lambda) $ if we
assume that 
\beq
F(\lambda)=\Theta (\lambda)\widetilde F(\lambda),\quad 
{\partial \over \partial \theta^\mu}\widetilde F(\lambda)=0,\quad 
\left(\partial_++\lambda^2\partial_-\right) \widetilde F=0.
\label{cond0}
\eeq 
\section{Combining the linear and the non linear equations}
So far we have solved them independently. We have derived the general solution of equation
\ref{dynud0} and a particular class of solutions of equations \ref{dynuu0}. Since the former is
linear, it should be satisfied oder by order in $g$ by the expansion \ref{pertdef}. At this moment
we are not able to do so beyond the zeroth order, which is already rather involved. In order that
$\Phi^{[0]}$ satisfies the self--consistency condition \ref{self8} it is sufficient that
$F(\lambda)$ satisfies them for any $\lambda$. Thus, according to equation \ref{solself8}, we should
be able to write 
$$
F(\lambda)=\left\{D_{\overline 1 \overline 2}-D_{\overline 3 \overline 4} +
 D_{\overline 5 \overline 6} -
D_{\overline 7 \overline 8}\right\}
\left(D_{\overline 5\overline 3}-D_{\overline 2\overline 8}\right)\left(D_{\overline 9}+
\partial^4_-\right)
\Psi(\lambda)=\Theta(\lambda) \widetilde F(\lambda)  
$$
This is achieved by letting 
$$
\widetilde F(\lambda)=\left\{\widetilde D_{\overline 1 \overline 2}-\widetilde D_{\overline 3
\overline 4} +
 \widetilde D_{\overline 5 \overline 6} -
\widetilde D_{\overline 7\overline 8}\right\}
\left(\widetilde D_{\overline 5\overline 3}-\widetilde D_{\overline 2\overline
8}\right)\left(\widetilde D_{\overline 9}+
\partial^4_-\right)
\widetilde  \Psi(\lambda)  
$$
where $\widetilde  \Psi(\lambda)=\Theta^{-1}(\lambda) \Psi(\lambda)$,  and we let
systematically 
$$
\widetilde D_{\overline \mu}= \Theta^{-1}(\lambda) D_{\overline \mu} \Theta(\lambda)
$$
Thus the question is whether we may choose $\widetilde \Psi$ such that equations \ref{cond0} hold. 
This is complicated but may be verified by expanding over $\lambda$. Rewrite schematically the
above as 
$$
\widetilde F(\lambda)= {\cal P}(\lambda) \widetilde  \Psi(\lambda) 
$$
At order zero, ${\cal P}^{[0]}(\lambda)$ is independent from $\theta$, so we simply impose that 
$$
\partial_{\theta^\mu}\Psi^{[0]}(\lambda)=\partial_+ \Psi^{[0]}(\lambda)=0. 
$$
Consider the order one. One has 
$$
{\cal P}^{[1]}=\sum_\mu \theta^{\mu }{\cal P}_\mu^{[1]}, \quad 
\partial_{\theta^\mu} {\cal P}_\mu^{[1]}=0. 
$$
Therefore at order one, we have 
$$
\widetilde F^{[1]}= {\cal P}^{[0]} \widetilde  \Psi^{[1]}+
\sum_\mu \theta^{\mu }{\cal P}_\mu^{[1]} \widetilde  \Psi^{[0]}
$$
Since  ${\cal P}^{[0]} $ is independent from $\theta$, we must have 
$$
\widetilde  \Psi^{[1]}=\sum_{\mu }\theta^{\mu }\Psi^{[1]}_\mu
$$
with 
$$
0={\cal P}^{[0]}  \widetilde  \Psi^{[1]}_\mu+
 {\cal P}_\mu^{[1]} \widetilde  \Psi^{[0]}
$$
The higher orders in $\lambda$ may be treated similarly. 

\section{Outlook}
It seems fair to say that the equation
$$
\sum_{\mu \overline \nu  }D_{\overline\nu}\left( {\bf R}^{-1}\, D_{\mu}\, {\bf R}\right)
\gamma^{ijk}_{{\mu \overline \nu  }}=0,
$$
$$
1\leq  i<j< k\leq 8. 
$$
has been the main obstacle in deriving classical solutions. We have been able to derive its
general solution by going to a special on--shell light cone gauge,  assuming  no dependence on
$x^i$, $i=1,\ldots, 8$. After this reduction we found  a striking analogy between the
other (non linear) equations and some of the  equations which arose from self--dual Yang Mills in
four bosonic variables.  In general there appear interesting novel structures (self--duality, Yang type 
equations) with superderivatives. They  seem to enjoy remarkable properties which, contrary to
their purely bosonic counterparts, are not restricted to four
variables. It will be probably 
 fruitful to push  this aspect further. 

There remains the consistency problem between the solutions of the  dynamical equations 
\ref{dynud0}, and \ref{dynuu0}. We have transformed the former into equations  
\ref{mix1}---\ref{mix3} and  the latter into equations \ref{nonleq}. After introducing $\Phi$ by
equation \ref{Phidef}, we have solved equations \ref{mix2}, \ref{mix3}, as well as equations 
\ref{nonleq}. Thus the difficulty  which remains is really the
consistency 
between equations \ref{mix1}
and the others.  At this time we are not able to go beyond the zeroth order in solving this problem.
We do not know whether the self--duality conditions may be imposed to the higher orders, but the
equations are not too promising.  Note that we have derived the
general solution of the linear part of the equations, but only a
subclass of solutions of the non-linear part. The situation may be improved either by deriving a more general
solution of the latter  (equations
\ref{nonleq}), or perhaps by only considering the $N\to \infty$ limit. This problem is left for the
future.  

    Finally,  the geometrical 
significance\cite{W86} of the flatness conditions \ref{flat}  is based on the super light--like lines
formulation of integrability along these lines. In other words, the construction of solutions to
ten dimensional supersymmetric Yang--Mills equations can be related, via the Radon--Penrose--type
transformation, to the construction of holomorphic bundles with some trivialisation on the space of
light--like lines. Various algebraic geometry constructions of the solutions to the problem, in
particular by the forward images method and the Baker functions were described in \cite{KM86}.
There, briefly speaking,  one deals with the following objects. Let $X$ be a complex supermanifold
(twistor superspace) of dimension $(17|8)$ parametrising super--light--like rays of dimension
$(1|8)$, and is endowed with some family of closed superspaces
$Y(u)\subset X$. 
These superspaces consist of $8$
dimensional quadrics with points $u\in U$, $U$ is a superdomain in ${\bf C}^{10|16}$
whose even part ${\bf C}^{10}$ is a complexification of a fundamental representation
of $O(1,9)$. Then locally free bundles on $X(U)$ being trivial under restriction on
$Y(u)$ correspond to the solutions of the ten dimensional supersymmetric Yang--Mills equations. 
It would be interesting to understand the geometrical meaning of the
present work along these lines.

\fin
\begin{thebibliography}{**}

\bibitem{W86}
E. Witten: Twistor--like transform in ten dimensions, Nuclear Physics 
{\bf B266} (1986), 245--264.

\bibitem{AFJ88}
E. Abdalla, M. Forger and M. Jacques: Higher conservation laws for 
ten--dimensional supersymmetric Yang--Mills theories, Nuclear Physics 
{\bf B307} (1988), 198--220.

\bibitem{CM89}
Ling--Lie Chau and B. Milewski: Linear systems and conservation laws in $d=10,
\, N=1$ supergravity, Physics Letters {\bf B216} (1989), 330--332.

\bibitem{N81}
B. E. W. Nilsson: Off--shell fields for the $10$-dimensional supersymmetric 
Yang--Mills theory, Nuclear Physics {\bf B188} (1981), 176.

\bibitem{BFSS96} 
T. Banks, W. Fischler, S.H. Shenker,  L. Susskind: 
M theory as a Matrix model: a conjecture, 
 hep-th/9610043, Phys. Rev. {\bf D55 } (1997) 5112. 


\bibitem{M97} L. Motl: Proposal on non perturbative superstring interactions, 
hep-th/9701025.

\bibitem{DVV97} R. Dijkgraff, E. Verlinde, H. Verlinde, 
Matrix string theory,  hep-th/9703030, Nucl. Phys. {\bf B500 } (1997) 43.
\bibitem{KM86}
M. M. Kapranov and Yu. I. Manin: Twistor transform and algebraic
geometry constructions of solutions to field theory equations, Russian 
Math. Surveys, {\bf 41:5} (1986), 85--107.

\bibitem{CDFN83}
E. Corrigan C. Devchand, D.B. Fairlie, and J. Nuyts: First order equations for gauge 
fields in spaces of dimensions greater than four Nucl. Phys.{\bf   B214} (1983) 452---464. 

\bibitem{LS89}
A. N. Leznov and M. V. Saveliev:
Exactly and Completely Integrable Nonlinear Dynamical Systems, 
Acta Applicandae Math. {\bf 16} (1989) 1--74. 
 
\end{thebibliography}
